\documentclass{IEEEtran}
\usepackage{amsmath}
\usepackage{amssymb}
\usepackage{graphicx}
\usepackage{comment}
\usepackage{xcolor}
\DeclareGraphicsExtensions{.pdf,.jpg,.png,.jpeg}

\usepackage[
style=ieee,doi=false,isbn=false,url=false,eprint=false,maxnames=4
]{biblatex}
\begin{document}
	
\title{{\normalsize Author preprint March 2026}\\
Quantifying resilience for distribution system customers with SALEDI }

\author{Arslan Ahmad 
\hspace{1.5cm} Ian Dobson 
\thanks{A. Ahmad and I.~Dobson are with Dept. Electrical \& Computer Engineering, Iowa State University, Ames Iowa USA; email: dobson@iastate.edu.
Support from USA NSF grants 2153163 and 2429602, Argonne National Laboratory, Iowa State University Electric Power Research Center, and PSerc project S110 is gratefully acknowledged.}}

\maketitle	

\begin{abstract}
The impact of routine smaller outages on distribution system customers in terms of customer minutes interrupted can be tracked using conventional reliability indices.
However, the customer minutes interrupted in large blackout events are extremely variable, and this makes it difficult to quantify the customer impact of these extreme events with resilience metrics. We solve this problem with the System Average Large Event Duration Index SALEDI that logarithmically transforms the customer minutes interrupted. We explain how this new resilience metric works, compare it with alternatives, quantify its statistical accuracy, and illustrate its practical use with standard outage data from five utilities.
\end{abstract}


\section{Introduction}

\looseness=-1
There has been a proliferation of resilience concepts and suggested metrics 
\cite{StankovicPS23, Pantelibook26,NanRESS17},
including advances describing and improving distribution system resilience \cite{DRWGreport,PaulRSER24,PoudelSJ20}.
However, tracking distribution system resilience with quantitative indices calculated from utility outage data, similarly to tracking reliability with the System Average Interruption Duration Index SAIDI, has remained elusive. In this paper we develop a novel {\bf S}ystem {\bf A}verage {\bf L}arge {\bf E}vent {\bf D}uration {\bf I}ndex SALEDI to quantify distribution system resilience in terms of customer minutes interrupted in large events. 

We write CMIp for 
{\bf C}ustomer {\bf M}inutes {\bf I}nterrupted {\bf p}er customer served. 

\noindent
Distinctive attributes of SALEDI are:
\begin{itemize}
\item SALEDI resolves the challenges of the rarity and high variability of CMIp in large resilience events by using a logarithmic transformation. This ensures that SALEDI can be practically calculated with reasonable accuracy from a few years of data.
\item SALEDI is complementary to SAIDI: SAIDI gives a stable indication of reliability by focusing on routine smaller outages and excluding major event days, whereas SALEDI gives a stable indication of resilience by focusing on large events and excluding small events.
\item SALEDI can be readily calculated from the same utility outage data used to calculate SAIDI.
\end{itemize}

Resilience addresses rare, high-impact extreme events, and these are inherently challenging to track with metrics because of their rarity and the high variability in the event magnitude. 
For example, it is well known that SAIDI, which describes annual reliability for distribution customers with CMIp, is erratic if major event days are not excluded from the calculation \cite{IEEE13662022,ChristiePD03}. 
More recently, experience with recorded distribution utility data shows that large event customer costs and customer minutes interrupted are highly variable over orders of magnitude \cite{AhmadPESL25,AhmadIET25,PandeyPESGM25}. 
The high variability is due to heavy tails in the distribution of event magnitudes.
The heavy tails and high variability do seem to be an inherent feature of extreme events in power systems\footnote{Power interrupted in transmission system blackouts is also well known to be heavy-tailed and highly variable over orders of magnitude \cite{CarrerasPS16}.}.

One consequence of the high variability is that some seemingly straightforward metrics calculated from real utility data that include extreme events are so erratic as to be impractical.
The impracticality manifests as these metrics requiring impractically large amounts of data to give a value that is not unduly fluctuating. 
To grapple with this problem, it is essential to check the statistical variability of any proposed resilience metrics, and this, while addressed in our recent initial work in \cite{AhmadPESL25,AhmadIET25}, has not seemed to be addressed at all in the rest of the literature on distribution system resilience. 
Accordingly, when this paper proposes SALEDI, we quantify its variability and demonstrate that it can be practically computed using only a few years of large-event utility data.

The paper makes the following contributions to calculating distribution system resilience from the CMIp of events:
\begin{itemize}
\item Uses outage data recorded by five utilities in the USA to demonstrate in practice the high variability and heavy-tailed distribution of event CMIp.
    \item Introduces SALEDI, a novel metric for distribution system resilience in terms of CMIp. 
    SALEDI essentially sums the logarithms of large event CMIp over a period of a few years.
    \item Factors SALEDI into the annual frequency of large events and another {\bf A}verage {\bf L}arge {\bf E}vent {\bf D}uration ALED metric. ALED is essentially the average logarithm of CMIp in large events.
    The factoring clarifies the relation between metrics that add and metrics that average the contributions from resilience events.
    \item Interprets SALEDI and ALED as areas under the tail of exceedance curves and relates them to logarithmic versions of Conditional Value at Risk CVAR. 
     Interprets SAIDI as the area under an exceedance curve to better link risk analysis methods with current industry practice.
    \item Shows that the logarithm is needed in the definitions of SALEDI and ALED by considering alternative metrics without the logarithm and showing that they require impractically large amounts of data. 
     \item Shows how ALED describes the tail of an exceedance function and thus the trend of large event CMIp.
    \item Demonstrates the practical calculation of SALEDI using standard outage data recorded by five utilities.
\end{itemize}

\subsection{Literature review}

There are many aspects of resilience not addressed in this paper, including transmission system resilience, organizational and community resilience, qualitative assessment, methods of hardening and faster restoration, the use of microgrids, backup generation, switching, and models of resilience processes and their optimization. 
And there are many good reviews of resilience, such as \cite{StankovicPS23,Pantelibook26, NanRESS17,IEEEwgs}, including reviews devoted to distribution system resilience \cite{DRWGreport,PaulRSER24}.
Therefore the following review is tailored to this paper by addressing relevant quantitative distribution system resilience indices.

\looseness=-1
Quantitative resilience metrics often describe the dimensions, slopes, and areas of a performance curve, which tracks a resilience measure over time over the course of a resilience event \cite{Pantelibook26,NanRESS17,
CarringtonPS21,IEEEwgs}. 
The metrics describe the number of outages, number of customers disconnected, outage and restore rates, durations, nadirs, and the area under the resilience curve. In this paper we are concerned with the area under a performance curve that tracks the number of customers out in an event, since this area is the same as the total customer minutes interrupted for all the outages in the event \cite{DobsonPESL23}.

A significant advance towards directly quantifying power system resilience is using Conditional Value at Risk (CVAR)\footnote{Alternative names for CVAR are average value-at-risk, expected shortfall, tail conditional expectation, mean expected loss, mean shortfall, and tail VAR.} to describe the risk of large events \cite{StrbacFTEES16,PoudelSJ20,Pantelibook26}. 
CVAR is the mean value of the large events, where the value can be energy not served or other measures. 
The importance of CVAR is that it directly describes the resilience risk of a top percentage of large events.
Moreover, the risk of large events measured by CVAR can be balanced with the reliability risk of smaller events by optimizing a linear combination of these risks. 
However, as shown in \cite{AhmadPESL25} for some distribution utility customer cost data and as suggested in \cite{CarrerasPS16} for transmission utility data, 
CVAR can be impractical to calculate from observed data because the high variability of blackout cost 
requires too many samples to get a reliable estimate.
The ALED metric of this paper shows how to solve this problem for CMIp with a logarithmic transformation.

For understanding and analyzing the heavy-tailed distributions encountered in our data, we rely especially on the excellent references \cite{Nairbook22,ClausetSIAM09}.

Heavy tails in financial data are addressed in \cite{Petersbook}, focusing mainly on heavy tails with exponent $\alpha>1$. 
A power law tail is approximated with a Hill estimator and used to estimate value at risk in \cite[Prop. 8.13]{Petersbook}.

\looseness=-1
There is some previous work using a logarithmic transformation for resilience metrics.
Pandey  \cite{PandeyPESGM25} considers the event customer hours divided by the number of customers interrupted and then takes a logarithm to define 
an Area Index of Resilience metric AIR. A Restoration Efficiency metric RE is the logarithm of the number of repair crews divided by the number of outages.
Adding RE and AIR gives a logarithmic REPAIR metric that combines both customer impact and repair efficiency.
The logarithm is justified for these metrics by the quantities varying over orders of magnitude for different events.

The customer risk metrics Annual Log Cost Resilience Index ALCRI and Average Log Event Cost ALEC are introduced in the letter 
 \cite{AhmadPESL25} and applied to one distribution system in \cite{AhmadIET25}.
ALCRI and ALEC estimate the customer cost from the CMIp, and use different scalings than SALEDI and ALED.

In this paper we modify, build on, and greatly expand the letter \cite{AhmadPESL25} by defining new logarithmic resilience metrics SALEDI and ALED to track CMIp rather than customer cost.
SALEDI and ALED are introduced, justified, interpreted, and compared to SAIDI and CVAR with extensive new material.

\section{SALEDI and SAIDI}

\subsection{SAIDI including major event days is highly variable}
\label{metrics}

It is well known for distribution system outage data that if major event days are not excluded, yearly SAIDI is an erratically variable metric of reliability \cite{IEEE13662022,ChristiePD03}.
This subsection sets up notation and reviews the variability of SAIDI.

Recall that sustained customer outages are those lasting more than 5 minutes.
Let $n_{\text{customer}}$ be the number of customers served by the utility.
Then the Customer Minutes Interrupted per served customer CMIp for the $i$th sustained outage is
\begin{align}
    m_i=\frac{\text{customer mins. interrupted in sustained outage $i$}}{n_{\text{customer}}}
    \label{normalizedcustomerminutes}
\end{align}

Over a period of one year, 
let $m_1,m_2,...,m_{n_{\rm outage}}$ be the CMIp for all the sustained outages in that year.
Then SAIDI including major event days for that year is defined as 
\begin{align}
    {\rm SAIDI}=\sum_{i=1}^{n_{\rm outage}} m_i
\end{align}

\begin{figure}[ht]
    \centering
    \includegraphics[width=1.0\linewidth]{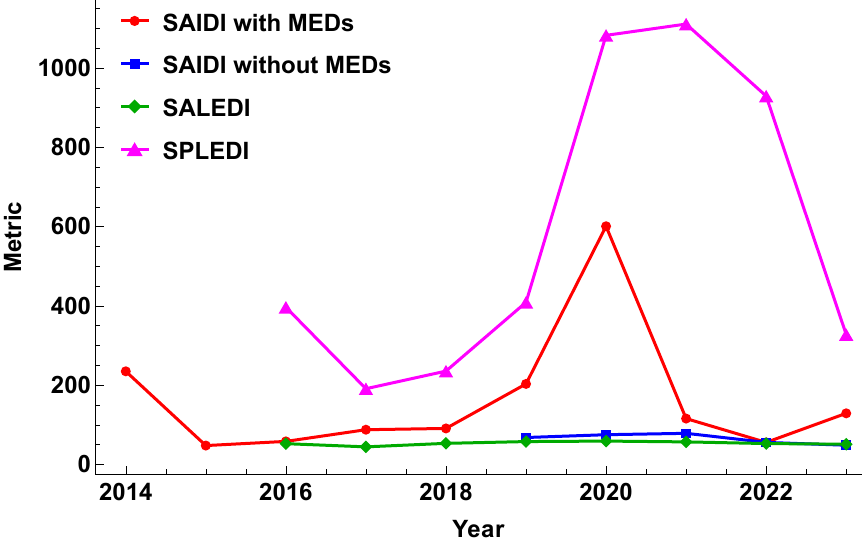}
    \caption{Annual fluctuations in SAIDI with and without major event days, and resilience indices SALEDI (with log) and SPLEDI (without log) for utility 4.}
    \label{fig:SaidiVsSaledi}
\end{figure}

When attempting to use SAIDI with major event days as a metric, its value often fluctuates significantly from year to year, as shown in Fig.~\ref{fig:SaidiVsSaledi}. 
The dominant source of fluctuation  is the random occurrence and variable magnitude of large events. That is,
SAIDI with major event days is responding more to the very high variability in the large events that can occur in each year than to the power system's underlying reliability or resilience.
After all, the distribution system reliability and resilience vary relatively slowly over many years as upgrades and changes in equipment and procedures are made.
We need resilience metrics to more stably track power system performance as it responds to weather extremes in order to guide mitigation and investment decisions.

For monitoring reliability, the problem of erratic SAIDI is traditionally solved by excluding the major event days from the annual SAIDI calculation \cite{IEEE13662022,ChristiePD03}, as illustrated in Fig.~\ref{fig:SaidiVsSaledi}.

\subsection{Utility outage data and grouping outages into events}

\looseness=-1
Distribution utilities typically record, automatically or manually, detailed outage data including the outage and restore times to the nearest minute and the number of customers interrupted.
The outage data for this paper are from
utility 1 (anonymous, Northeast USA),
utility 2 (Muscatine Power and Water, Iowa, USA), 
and the following parts of Massachusetts, USA:
utility 3 (north central MA, Middlesex and Worcester),
utility 4 (western MA counties, Berkshire, Franklin, Hampden, Hampshire), 
utility 5 (mainly Suffolk County and some neighboring areas from Middlesex and Norfolk counties).
We gratefully acknowledge Muscatine Power and Water for generously providing the utility 2 outage data, and the Commonwealth of Massachusetts for making the utility 3,4,5 outage data publicly available at \cite{massData}.

For a resilience analysis, we extract from these data events in which outages bunch up and overlap \cite{CarringtonPS21,StankovicPS23}.
The events range in size from isolated single outages to large events with hundreds of outages caused by extreme weather. 

In this paper, resilience events are defined as a group of outages that overlap in time\footnote{
Events can also be defined as a group of outages that overlap both in time and are nearby in location, which is especially useful when the utility serves a large geographic area \cite{ahmadArxiv25}.}. 
Time-based grouping of outages into events works as follows: An event starts when an initial outage starts. All the subsequent outages that start before all the previous outages are restored are grouped into the same event, while assuming that each outage duration is limited to 3 hours\footnote{The 3-hour limitation is used only for grouping purposes, and the actual duration of each outage is retained and used in all subsequent calculations. The 3-hour limitation ensures that a few very long outages do not cause the grouping of unrelated outages in the same event.}.
More details of event definition are given in \cite{ahmadArxiv25}.

\subsection{Event customer minutes interrupted and large events} 
Once events are defined, the CMIp $M_i$ for event~$i$ is simply the sum of the CMIp of the sustained outages in event $i$:
\begin{align}
    M_i=\sum_{\text{outages $j$ in event $i$}}m_j
    \label{M}
\end{align}
$M_i$ is the same as SAIDI that is evaluated only for the outages in event $i$.
The observed large event CMIp varies over several orders of magnitude as can be seen in Fig.~\ref{fig:exceedancetail}. 

For resilience we focus on large events by setting a threshold $M_{\rm large}$ so that the large events are the events with CMIp $M\ge M_{\rm large}$. 
Section \ref{threshold} explains how 
to choose a suitable value of
$M_{\rm large}$.

\subsection{SALEDI definition}

Over a period of $n_{\rm year}$ years, 
let $M_1,M_2,...,M_{n_{\rm large}}$ be the CMIp for all the large events in that period. 
The number of years $n_{\rm year}$ can be one if there are a sufficient number of large events in one year; otherwise, we take $n_{\rm year}$ to be more than one year to accumulate enough large events as detailed in section~\ref{sec:PracticalDetails}.

We define the System Average Large Event Duration Index
\begin{align}
    \text{SALEDI}=\frac{1}{n_{\rm year}}
    \sum_{i=1}^{n_{\rm large}}\ln \frac{M_i}{M_{\rm large}}
    \label{SALEDI}
\end{align}
Note the simplicity of SALEDI: First add up over a period of $n_{\rm year}$ years the logarithm of the CMIp in the large events normalized by $M_{\rm large}$. Then scale to an annual index by dividing by $n_{\rm year}$.
SALEDI is a system-average metric that assesses impact per customer served, because $m_i$ and $M_i$ are normalized by the number of customers served in (\ref{normalizedcustomerminutes}).

\subsection{SALEDI incorporates large event frequency \& magnitude}
\looseness=-1
SALEDI defined by \eqref{SALEDI} responds to both the frequency and magnitude of large events: If large events are more frequent, $n_{\rm large}$ and SALEDI increase. If large events have more CMIp, SALEDI increases.
To show this more precisely,
rewrite (\ref{SALEDI}) as
\begin{align}
    \text{SALEDI}&=
    \left(
    \frac{n_{\rm large}}{n_{\rm year}}
    \right)
    \left(
    \frac{1}{n_{\rm large}}
    \sum_{i=1}^{n_{\rm large}}\ln \frac{M_i}{M_{\rm large}}
    \right)\notag\\
    &=f_{\rm large}\, {\rm ALED},
    \label{SALEDIfactor}\\[2mm]
\text{where  }\qquad f_{\rm large}&= 
n_{\rm large}/n_{\rm year}
\label{freq}
\end{align} 
is the annual frequency of large events estimated over $n_{\rm year}$ years, and the Average Large Event Duration metric
\begin{align} 
{\rm ALED}=\frac{1}{n_{\rm large}}
    \sum_{i=1}^{n_{\rm large}}\ln \frac{M_i}{M_{\rm large}}
    \label{ALED}
\end{align}
is the logarithm of the event CMIp normalized by $M_{\rm large}$ averaged over large events. 
It is clear from (\ref{SALEDIfactor}) that SALEDI is proportional to 
 the annual frequency of large events (\ref{freq}) as well as the average log normalized CMIp of large events ALED (\ref{ALED}).

Since SALEDI sums over events, SALEDI can be usefully decomposed by event cause. For example, SALEDI (total) = SALEDI (caused by trees) +
SALEDI (not caused by trees). This is useful in estimating the effectiveness of mitigation  addressing a specific  cause (such as tree trimming).

ALED and SALEDI are responsive to resilience investments.
In particular, a 10\% decrease in CMIp of all large events leads to a change of $\ln 0.9=-0.105$ in ALED and a change of $-0.105 f_{\rm large}$ in SALEDI (the change can be a larger decrease because any large events with CMIp less than $1.1 M_{\rm large}$ are no longer large events when their CMIp decrease by 10\%). A 10\% decrease in all large event restoration times also leads to a $-0.105$ change in ALED and a $-0.105 f_{\rm large}$ change in 
SALEDI. A 10\% increase in the frequency of large events increases SALEDI by approximately $\ln 1.1=0.095$ but ALED is approximately unchanged.

\subsection{Results}

\begin{figure}[ht]
    \centering
    \includegraphics[width=1.0\linewidth]{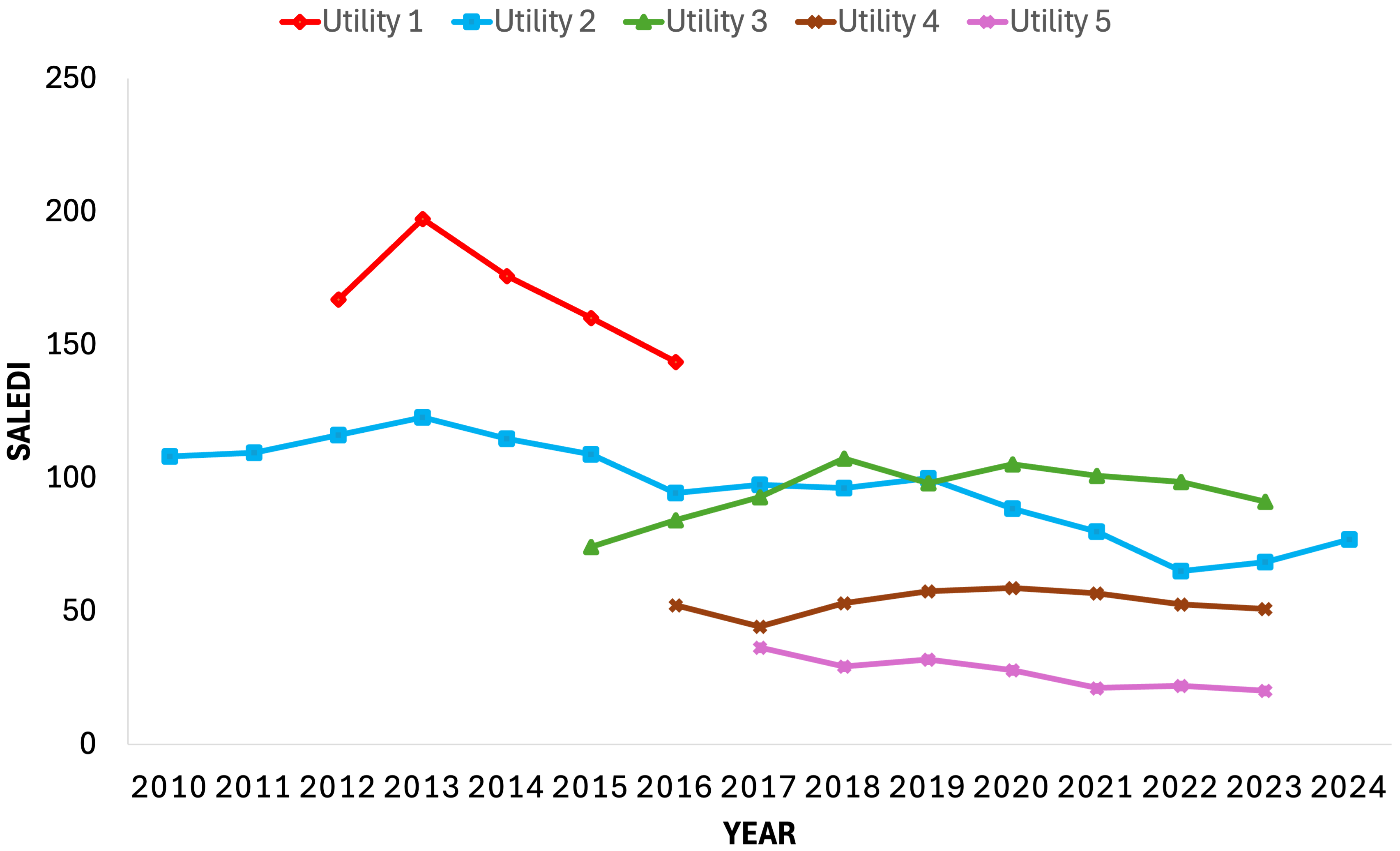}
    \caption{Tracking SALEDI over years. For each utility, a sliding window of duration $n_{\rm year}^{\rm min}$ rounded to an integer is used.}
    \label{fig:salediOfAllUtilities}
\end{figure}

Table \ref{utilitydata} describes the utility data, calculated quantities, and the values of SALEDI and ALED for the complete data of the last $n_{\rm year}^{\rm min}$ years of each utility, where $n_{\rm year}^{\rm min}$ is determined to ensure that SALEDI's relative standard error $\rm RSE_{SAL} = 0.1$ as discussed in section~\ref{sec:PracticalDetails}.
Fig.~\ref{fig:salediOfAllUtilities} shows how SALEDI varies when calculated over a sliding window of length $n_{\rm year}^{\rm min}$ rounded to an integer number of years. 
Indeed, Fig.~\ref{fig:salediOfAllUtilities} shows SALEDI describing the power system resilience with respect to large event customer impacts with low variability.
The low variability of SALEDI is explained and quantified in section \ref{theory}.
The only heavily urbanized case we consider is utility 5, and this might explain its lower SALEDI and ALED and higher resilience.

\begin{table}[hbpt]
	\caption{Utility data and metrics} 
	\label{utilitydata}
	\centering
\begin{tabular}{ l @{\hspace{-1pt}} c c c c c}
                        &Utility-1  &Utility-2  &Utility-3  &Utility-4  &Utility-5  \\ \hline
${n_{\rm allevent}}$        & 5712      & 2532      & 3783      & 6919      & 6212     \\
${n_{\rm year}^{\rm all}}$  & 6         & 17.4      & 11        & 10        & 11     \\
$f_{\rm large}^{\rm all}$   & 133.6     & 73.5      & 65.5      & 57.1      & 40.4     \\
$M_{\rm maxobs}$            & 820.4    & 287.9    & 654.4    & 344.9    & 15.6     \\
$n_{\rm customer}$         & 307791   & 11612    & 30743    & 193185   & 483970    \\[2mm]

$M_{\rm large}$             & 0.114     & 0.021     & 0.172     & 0.253     & 0.303     \\
Quantile $q$                & 0.86      & 0.50      & 0.81      & 0.92      & 0.93     \\
$n_{\rm year}^{\rm min}$              & 1.50         & 2.72        & 3.05         & 3.50         & 4.96    \\
$\alpha$                    & 0.83      & 0.74      & 0.71      & 1.07      & 1.44     \\[2 mm]

${\rm RSE}_{{Pb}}$      & 40.06     & 60.96     & 25.56     & 33.68     & 13.44    \\
${\rm RSE}_{LNb}$       & 39.48     & 64.94     & 21.44     & 8.74      & 2.66  \\
$n_{{\rm large}\,Pb}^{\rm minSPL}$&$1.6\mathrm{E}{5}$&$3.7\mathrm{E}{5}$&$6.5\mathrm{E}{4}$&$1.1\mathrm{E}{5}$&$1.8\mathrm{E}{4}$\\
$n_{{\rm large}LNb}^{\rm minSPL}$&$1.6\mathrm{E}{5}$&$4.2\mathrm{E}{5}$&$4.6\mathrm{E}{4}$&$7700$&$800$\\
[2 mm]

\multicolumn{6}{c}{following rows calculated from the last $n_{\rm year}^{\rm min}$ years of data so that  }\\
\multicolumn{6}{c}{$n_{\rm large}=n_{\rm large}^{\rm min}=200$, $\rm RSE_{SAL}=0.1$, and RSE$_{\rm ALED}=0.07$}\\[1mm]
$f_{\rm large}$             & 133.6    & 73.5      & 65.5     & 57.1    & 40.4    \\
SALEDI                      & 134.8     & 111.      & 93.7     & 55.9      & 26.7    \\
ALED                        & 1.01      & 1.51      & 1.43      & 0.98      & 0.66  \\

\hline\\[-2.2mm]

\end{tabular}
\end{table}

\section{Technical basis of SALEDI}
\label{theory}
This section develops and explains the theory behind SALEDI, calculates its statistical variation, and relates it to the area under an  exceedance function.

\subsection{Observed exceedance functions}

\begin{figure}[ht]
    \centering
    \includegraphics[width=1.0\linewidth]{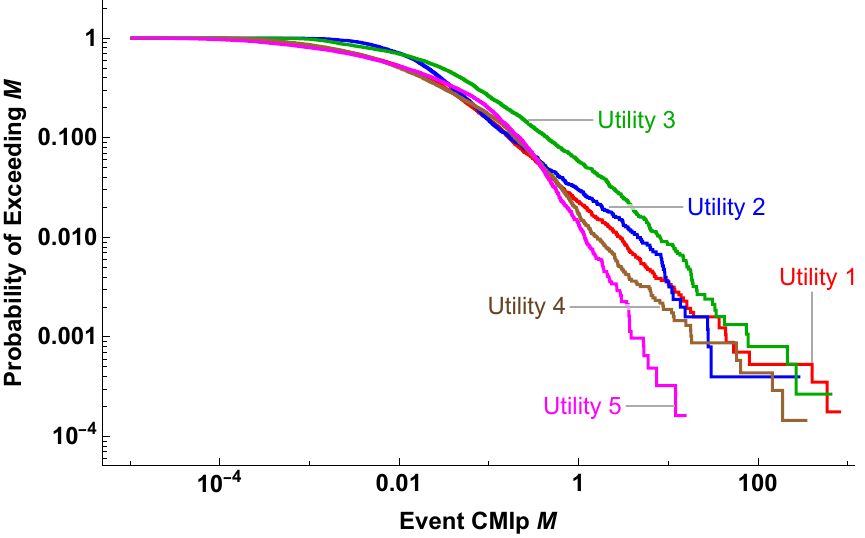}
    \caption{\!\!\!Exceedance functions $\overline{F}(M)$ for all events of utilities 1-5; log-log plot. }
    \label{fig:exceedance}
\end{figure}

The CMIp for all the events in the observed data are $M_1^{\rm all},M_2^{\rm all},...,M_{n_{\rm allevent}}^{\rm all}$.
Consider the probability exceedance function\footnote{$\overline{F}(M)$ is also the empirical CCDF 
of the event CMIp.}
\begin{align}
    \overline{F}(M)&=\text{fraction of all event CMIp}>M\notag\\
    &= \frac{1}{n_{\rm allevent}}\sum_{i=1}^{n_{\rm allevent}}I[M_i^{\rm all}>M]
    \label{FM}
\end{align}
where $I$ is the indicator function.
 $\overline{F}(M)$ describes risk in the sense that it combines event frequency and event impact in terms of CMIp.
Fig.~\ref{fig:exceedance} shows the exceedance functions $\overline{F}(M)$ for the five utilities on a log-log plot. 
For quantifying resilience, we are interested in characterizing the tail $M\ge M_{\rm large}$ of the exceedance function. 
We choose an $M_{\rm large}$ such that the tail $M\ge M_{\rm large}$ has an approximate straight-line behavior on the log-log plot.
The slope magnitude of the tail is denoted by $\alpha$.
$M_{\rm large}$ and $\alpha$ are shown in Table~\ref{utilitydata}.
More discussion on the selection of $M_{\rm large}$ is in section~\ref{sec:ChoosingMlarge}.

\subsection{Linear approximation to the exceedance function  tail}

\begin{figure}[ht]
    \centering
    \includegraphics[width=1.0\linewidth]{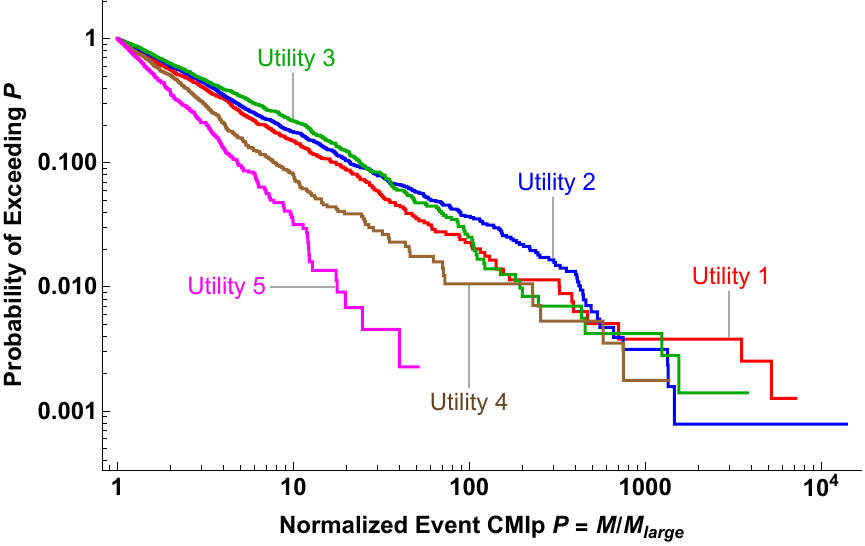}
    \caption{Tails of the exceedance functions of Fig.~\ref{fig:exceedance} with normalized event CMIp $P=M/M_{\rm large}$ on the horizontal axis;  log-log plot. }
    \label{fig:exceedancetail}
\end{figure}

Since we are interested in characterizing the tail of the exceedance function, we select only the tail data $M\ge M_{\rm large}$, divide it by $M_{\rm large}$ to obtain the normalized CMIp $P=M/M_{\rm large}$, and plot it on a log-log scale as shown in Fig.~\ref{fig:exceedancetail}.
The tails in Fig.~\ref{fig:exceedancetail} are approximately straight lines, and therefore have approximately the  power-law behavior that is described by a Pareto distribution $P$ with slope magnitude $\alpha$:
\begin{align}
    \overline{F}_P(p)=
   p^{-\alpha}
    ,\qquad p\ge 1,~\alpha>0
    \label{powerlaw}
\end{align}
To verify the straight-line behavior of (\ref{powerlaw}) on a log-log plot, take
the logarithm of (\ref{powerlaw}) to obtain $\ln{\overline F}(p)=-\alpha\ln p$.

Formal hypothesis testing for the tails for all five utilities indicates that 
linear approximations to the tails in the log-log plots are reasonable; see Appendix~\ref{AppendixA} for details.

\subsection{Log transformation converts heavy tail to light tail, and estimating the slope magnitude $\alpha$ of the heavy tail}

\begin{figure}[ht]
    \centering
    \includegraphics[width=1.0\linewidth]{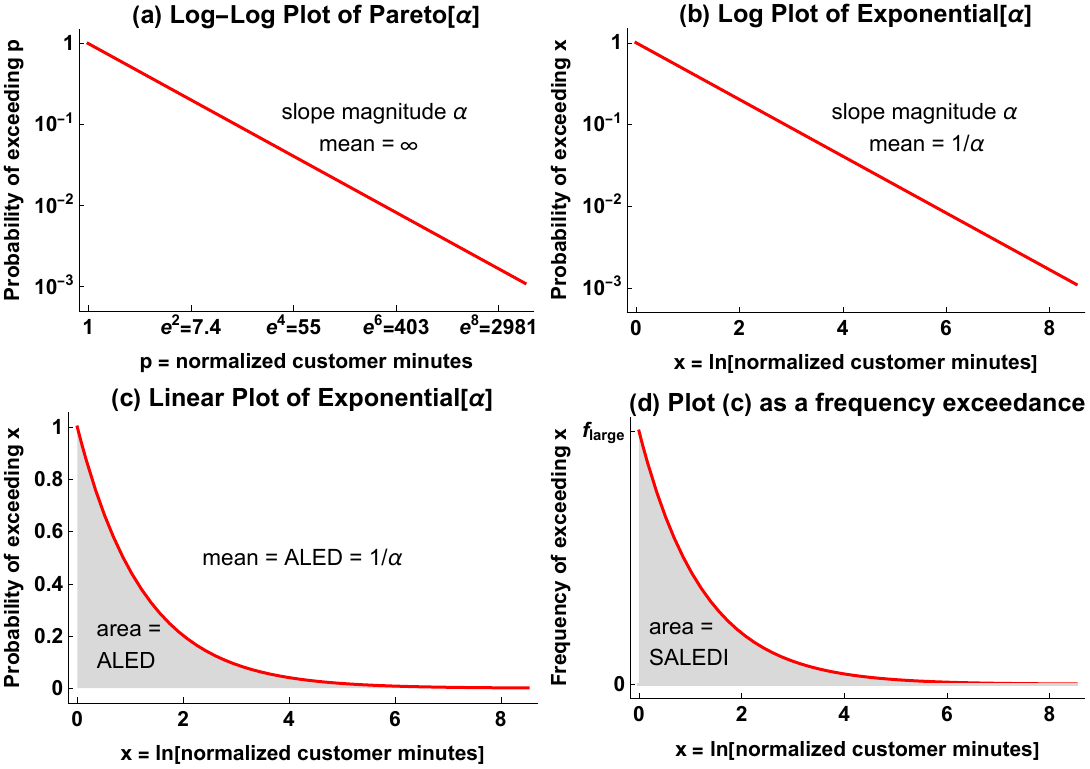}
    \caption{Plot (a) shows on a log-log plot the idealized Pareto probability exceedance function \eqref{powerlaw} of the normalized CMIp $M/M_{\rm large}$ for the large event tail. The slope magnitude $\alpha$ characterizes the tail. In this plot $\alpha=0.8<1$ so that the mean ${\rm E}P$ is infinite.
    Plot (b) applies a logarithm to the same normalized CMIp by relabeling the horizontal axis. Since plot (b) is an exceedance function that is the same straight line of slope magnitude $\alpha$ but now on a log plot, the log-transformed data $X=\ln P$ is an exponential distribution of rate $\alpha$ and finite mean $1/\alpha$. Plot (c) is the same as plot (b) but now on a linear plot. The area under plot (c) is the ALED metric. Plot (d) is the same as plot (c) except that the vertical axis is rescaled to show a frequency exceedance function with annual large event frequency $f_{\rm large}$. The area under plot (d) is SALEDI = $f_{\rm large}$ ALED.}
    \label{fig:logplots}
\end{figure}

If we take log of the tail data and define $X=\ln P$, then $X$ has an exponential distribution with rate parameter $\alpha$:
\begin{align}
\overline{F}_X(x)
    &=e^{-\alpha x},\quad x\ge 0
    \label{exp}
\end{align} 
This follows since 
$\overline{F}_X(x)={\rm P}[X>x]={\rm P}[\ln P>x]={\rm P}[P>e^x]
=\overline{F}_P(e^x)=
    e^{-\alpha x}$. 
That is, the logarithm of the heavy-tailed Pareto data with slope magnitude $\alpha$ has a light-tailed exponential distribution with rate $\alpha$; also see Fig. \ref{fig:logplots}. 

The linear approximation of the power-law tail on a log-log plot has slope magnitude $\alpha$. 
Equation (\ref{ALED}) shows that ${\rm ALED}$ estimates the mean of the exponential distribution (\ref{exp}), which is $\alpha^{-1}$. 
Therefore ${\rm ALED}^{-1}$
estimates the slope magnitude $\alpha$ of the power law tail.
Indeed, ${\rm ALED}^{-1}$ 
is the Hill estimator\footnote{
Determining a suitable value of $M_{\rm large}$ is a well-known delicate
issue for the Hill estimator.} for $\alpha$ \cite{Nairbook22,Resnickbook07}.
This important relation between ALED and the slope magnitude $\alpha$ shows how ALED describes the power law tail 
of direct interest to resilience risk.
Moreover, the power law slope magnitude captures the linear trend of the large event CMIp, and this trend governs the largest blackouts that may be experienced as discussed in subsection \ref{extrapolate}. Since ${\rm ALED}$ is a factor of ${\rm SALEDI}$ as shown in (\ref{SALEDIfactor}), ${\rm SALEDI}$ also incorporates the linear trend of the large event CMIp.

\subsection{Statistical variation of SALEDI}
\label{statSALEDI}
We now estimate the statistical variation of SALEDI with its Relative Standard Error ${\rm RSE}_{\rm SAL}$,
which is the standard deviation of SALEDI divided by the mean of SALEDI. 
The statistical variation of SALEDI has two sources: the number of large events and the magnitude of large events.

To estimate the statistical variation of the number of large events,
let $N_{\rm large}$ be the random variable that is the number of large events in $n_{\rm year}$ years of data. 
Then $N_{\rm large}$ is modeled as a Poisson[$ n_{\rm large}$] random variable.

For the statistical variation in the magnitude of large events, we approximate the tail of the distribution with a straight line on the log-log plot so that it follows the Pareto distribution (\ref{powerlaw}).
Then SALEDI can be considered to be a random variable that is the sum of independent and identically distributed samples $X_1$, $X_2$, ..., $X_{N_{\rm large}}$ from the 
log transformed data (\ref{exp}) divided by $n_{\rm year}$: 
\begin{align}
    {\rm SALEDI}=\frac{1}{n_{\rm year}}
    \sum_{i=1}^{N_{\rm large}}X_i
    \label{SALEDIrv}
\end{align}
Since $X$ follows the exponential distribution (\ref{exp})\footnote{Upper bounding (\ref{exp}) with the maximum normalized CMIp $M_{\rm max}/M_{\rm large}$ considered in subsection~\ref{whylog} makes no practical difference.},
 ${\rm E}X=\alpha^{-1}$ and ${\rm Var}X=\alpha^{-2}$ so that ${\rm RSE}_X=\sigma[X]/{\rm E}X=1$. Then Appendix \ref{AppendixB} calculates
the RSE of SALEDI and ALED as
\begin{align}
    {\rm RSE}_{\rm SAL} =\frac{\sqrt{1+({\rm RSE}_X)^2}}{\sqrt{ n_{\rm large}}}
    =\frac{\sqrt{2}}{\sqrt{ n_{\rm large}}}
    \label{RSESALEDI}\\
{\rm RSE}_{\rm ALED}=
\frac{{\rm RSE}_X}{\sqrt{ n_{\rm large}}}
    =\frac{1}{\sqrt{ n_{\rm large}}}
    \label{RSEALED}
\end{align}

\subsection{Exceedance function area interpretations}

This section relates SAIDI and SALEDI to areas under exceedance functions. Since exceedance functions are used for risk analysis \cite{KaplanRA81}, this relates SAIDI and SALEDI to methods of risk analysis.

Consider a general empirical frequency exceedance function for positive data $x_1,x_2,...,x_k$:
\begin{align}
    \overline G(x)&=\text{number of data points}>x
    = \sum_{i=1}^{k}I[x_i>x]
    \label{Gdefgeneral}
\end{align}
Then by taking horizontal slices of the area under $\overline G$,
\begin{align}
    \int_0^\infty \!\!\!\!\overline G(x)dx&=
    \sum_{i=1}^{k} [(k-i+1)-(k-i)]x_i=\sum_{i=1}^{k}x_i
    \label{Gintegral}
\end{align}
A scaled version of (\ref{Gintegral}) for a 
general empirical probability exceedance function $\overline F(x)=\frac{1}{k}\overline G(x)$
is\footnote{Equation (\ref{Fintegral}) is the empirical distribution case of the standard result that the mean of a nonnegative random variable is the integral of its probability exceedance function.}
\begin{align}
    \int_0^\infty \!\!\!\!\overline F(x)dx=\frac{1}{k}\sum_{i=1}^{k}x_i={\rm Mean}\{x_1,x_2,...,x_k\}
    \label{Fintegral}
\end{align}

\subsubsection{Area interpretation for SAIDI}
\begin{figure}[h] \centering\includegraphics[width=1.0\linewidth]{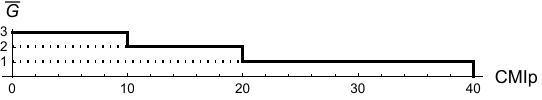}
    \caption{Simple example of 3 outages with CMIp values 10, 20, 40 respectively. Area under exceedance curve $\overline G = 10+20+40= {\rm SAIDI}$ for the 3 outages.}
    \label{fig:SAIDIarea}
\end{figure}

Let $m_1,m_2,...,m_k$
be the CMIp in each outage, and
consider the frequency exceedance function
\begin{align}
    \overline G(m)&=\text{number of outage CMIp}>m
    = \sum_{i=1}^{k}I[m_i>m]\notag
\end{align}
Then, using (\ref{Gintegral}), SAIDI 
is the area under the frequency exceedance function (see example in Fig~\ref{fig:SAIDIarea}):
\begin{align}
    \int_0^\infty \!\!\!\!\overline G(m)dm&=\sum_{i=1}^{k}m_i={\rm SAIDI}
    \label{G}
\end{align}
Also $\overline G(0)=k=$ number of outages.

\subsubsection{Area interpretation for SALEDI and ALED}
Let $M_1$,$M_2$,...,$M_{n_{\rm large}}$ be the
CMIp for large events, and
consider the annual frequency exceedance function of the tail 
\begin{align}
    \overline G_S(L)
    &= \frac{1}{n_{\rm year}}\sum_{i=1}^{n_{\rm large}}I[\ln \frac{M_i}{M_{\rm large}}>L]
\end{align}
Using (\ref{Gintegral}),
SALEDI is the area under $\overline G_S$ (see Fig. \ref{fig:logplots}(d)):
\begin{align}
    \int_{0}^\infty&\!\! \overline G_S(L)dL
    =\frac{1}{n_{\rm year}}\sum_{i=1}^{n_{\rm large}}\ln \frac{M_i}{M_{\rm large}}={\rm SALEDI}
\end{align}
Moreover, 
$
\overline G_S(0)=n_{\rm large}/{n_{\rm year}}=f_{\rm large}
$.

Similarly, consider the probability exceedance function for the tail 
$\overline F_S(L)= \frac{n_{\rm year}}{n_{\rm large}} G_S(L)$.
Then, using (\ref{Fintegral}) and as shown in Fig. \ref{fig:logplots}(c),
ALED is the area under $\overline F_S$:
\begin{align}
    \int_{0}^\infty&\!\! \overline F_S(L)dL=\frac{1}{n_{\rm large}}\sum_{i=1}^{n_{\rm large}}\ln \frac{M_i}{M_{\rm large}}={\rm ALED}
\end{align}

\section{Comparing SALEDI with alternatives}

This section 1) explains why the logarithm in SALEDI and ALED is necessary by describing the high variability when the logarithm is removed, 2) compares SALEDI and ALED with SAIDI and CVAR, and 3) discusses how ALED describes the trend governing the largest blackouts.

\subsection{Why the logarithm in SALEDI is needed}
\label{whylog}
Consider a Seemingly Plausible Large Event Duration Index SPLEDI that is SALEDI without the logarithm:
\begin{align}
 {\rm SPLEDI}=\frac{1}{n_{\rm year}}
    \sum_{i=1}^{n_{\rm large}}\frac{M_i}{M_{\rm large}}
    \label{SPLEDI}
\end{align} 
SPLEDI adds the normalized CMIp of the large events over $n_{\rm year}$ years\footnote{Due to (\ref{M}), adding the normalized large event CMIp to obtain SPLEDI in (\ref{SPLEDI}) is the same as adding the normalized CMIp of all the outages in the large events.} and then divides by $n_{\rm year}$ to obtain an annual index.
Also, factoring similarly to (\ref{SALEDIfactor}),
\begin{align}
 {\rm SPLEDI}&= 
    f_{\rm large}\, {\rm SPALED}
\label{SPLEDIlargefactor}\\
\text{where    }\qquad
{\rm SPALED}&=\frac{1}{n_{\rm large}}
    \sum_{i=1}^{n_{\rm large}}\frac{M_i}{M_{\rm large}}
    \label{SPALED}
\end{align}
is a Seemingly Plausible Average Large Event Duration Index that is the mean number of normalized CMIp for large events. SPALED is ALED without the logarithm.
We now quantify the variability of SPLEDI and SPALED to show that practical estimates of 
SPLEDI and SPALED require too many samples.

If we assume that the normalized CMIp in the tail have the Pareto distribution $P$ in (\ref{powerlaw}), then it is clear that SPLEDI and SPALED are impractical and excessively variable for the smaller values of absolute slope magnitude $\alpha$.
If $\alpha\le 1$, the mean ${\rm E}P$ is infinite, and the strong law of large numbers
\cite[section 7.5]{Grimmettbook} 
implies that estimates of the mean such as SPALED that add up samples and divide by the number of samples $n_{\rm large}$ do not converge to anything as $n_{\rm large}$ increases.
Since SPALED is a factor of SPLEDI as shown in (\ref{SPLEDIlargefactor}), SPLEDI also does not converge. $\alpha\le 1$ is observed for utilities 1,2,3 in our data, and $\alpha=1.07$ for utility 4 and $\alpha=1.44$ for utility 5.
If $1<\alpha<2$, both the standard deviation $\sigma(P)$ and the standard deviation of estimates of the mean 
$\sigma(P)/\sqrt{n_{\rm large}}$ are infinite, and we can expect that although SPALED and SPLEDI will converge in theory to a value as $n_{\rm large}$ increases, the convergence will be so slow that it will require a very large number of events $n_{\rm large}$ for any usable accuracy.

However, the Pareto distribution (\ref{powerlaw}) has an unbounded tail, indicating no limit to the blackout magnitude.
Since this is not realistic, we need to calculate the variability of SPLEDI and SPALED with distributions that assume a maximum $M_{\rm max}$ of CMIp.
There is some uncertainty about such a maximum possible blackout. To obtain a rough estimate, we assume that the maximum possible blackout with $M_{\rm max} = 43\,830$ minutes per customer corresponds to a blackout affecting all customers for one month,
or equivalently,
a blackout of half the customers for two months.

We fit the observed data with two different distributions, each limited by the maximum blackout CMIp $M_{\rm max}$. Since these distributions are bounded, they have large but finite means and variances, and estimates of their means do eventually converge. The issue is the slow rate of convergence and the large number of samples needed for usable statistical variability of the mean, as explained below.

A variant $P_b$ of the Pareto distribution (\ref{powerlaw}) with upper bound $p_{\rm max}=M_{\rm max}/M_{\rm large}$ is given by the exceedance function
\begin{align}
\overline{F}_{\!P_b}(p)=
\frac{p^{-\alpha} -p_{\rm max}^{-\alpha}}{
1-p_{\rm max}^{-\alpha}},\quad 1\le p\le p_{\rm max}
\label{Pb}
\end{align}
A bounded variant $LN_{b}$ of the lognormal distribution conditions the standard lognormal distribution 
to have $1 \le p\leq p_{\rm max}$.
That is, $LN_{b}$ has probability exceedance function
\begin{align}
\overline{F}_{\!LN_b}(p)=&
\frac{\overline{F}_N(\ln p) -\overline{F}_N(\ln p_{\rm max})}{
\overline{F}_N(0) -\overline{F}_N(\ln p_{\rm max})},\quad 1\le p\le p_{\rm max}
\end{align}
where $\overline{F}_N$ is the probability exceedance function of a normal distribution with mean $\mu$ and standard deviation $\sigma$.

Let $Y$ be the random variable for the normalized event CMIp $M/M_{\rm large}$. We will take $Y\sim P_b$ or $Y\sim LN_{b}$.
According to (\ref{SPALED}) and (\ref{SPLEDI}), SPALED is the mean of $N_{\rm large}$ samples of $Y$, where $N_{\rm large}$ has mean $ n_{\rm large}$, and SPLEDI is SPALED multiplied by annual frequency $f_{\rm large}$.
Then, using Appendix \ref{AppendixB}, the relative standard error of SPLEDI is
\begin{align}
    {\rm RSE}_{\rm SPL}
   &=\frac{\sqrt{1+({\rm RSE}_Y)^2}}{\sqrt{ n_{\rm large}}}
    \label{RSESPLEDI}
\end{align}
Then, similarly to section \ref{accuracy}, where we take  ${\rm RSE}_{\rm SAL}=0.1$, the minimum number of large events to have ${\rm RSE}_{\rm SPL}=0.1$ is
\begin{align}
n_{{\rm large}\,Y}^{\rm minSPL}=
100(1+({\rm RSE}_Y)^2)
\end{align}

For approximating the bounded distribution of CMIp with $Pb$, Table~\ref{utilitydata} shows that utilities 1,2,3,4 have $n_{{\rm large}\,Pb}^{\rm minSPL}$ at least two orders of magnitude greater than $n_{\rm large}^{\rm min}=200$. 
That is, for the same statistical accuracy, SPLEDI requires at least two orders of magnitude more large events than SALEDI, making it impractical as a resilience metric calculated from observed data. For example, if SALEDI requires 2 years of data to accumulate enough large events for reasonable statistical accuracy, then SPLEDI requires  at least 200 years.
Note that the ratio $n_{{\rm large}\,Pb}^{\rm minSPL}/n_{\rm large}^{\rm min}$ is $[1+({\rm RSE}_{Pb})^2]/2$, which is independent of the choice of statistical accuracy ${\rm RSE}_{\rm SAL}=0.1$.
For utility 5,
$n_{{\rm large}\,Pb}^{\rm minSPL}= 18\,000$  which is almost two order of magnitudes greater than $n_{\rm large}^{\rm min}=200$ and so requires more large events by a factor of $90$.

For approximating the bounded distribution of CMIp with $LNb$, Table~\ref{utilitydata} shows that utilities 1,2,3 have $n_{{\rm large}LNb}^{\rm minSPL}$ two orders of magnitude greater than $n_{\rm large}^{\rm min}=200$, so that SPLEDI requires at least two orders of magnitude more large events than SALEDI for the same accuracy.
Utility 4 requires more large events by a factor of $38$, and utility 5 requires more large events by a factor of $4$.

Similarly, using Appendix \ref{AppendixB}, the relative standard error of SPALED is
$    {\rm RSE}_{\rm SPA}
   ={\rm RSE}_Y/\sqrt{ n_{\rm large}}$
so that the minimum number of large events to have ${\rm RSE}_{\rm SPA}=0.1$ is
$n_{{\rm large}\,Y}^{\rm minSPA}=
100({\rm RSE}_Y)^2$, and the ratio $n_{{\rm large}\,Y}^{\rm minSPA}/n_{\rm large}^{\rm min}=({\rm RSE}_Y)^2$.
Then $n_{{\rm large}\,Y}^{\rm minSPA}$ for both $Y=Pb$ and $Y=LNb$ is two orders of magnitude larger than ALED's $n_{\rm large}^{\rm minALE}=100$ (obtained starting from \eqref{RSEALED})
 except that $n_{{\rm large}LNb}^{\rm minSPA}$ = 7600 and 700 for utilities 4 and 5 respectively.
Even if it could in theory be computed from large quantities of data,  SPALED, since it is  a  mean, cannot be interpreted as a typical value due to the heavy tails.

Overall, we conclude that, with the possible exception of utility~5, calculation of SPLEDI or SPALED requires impractically more large events and years of data to be observed than SALEDI or ALED. That is, the logarithmic transformation is necessary for practical metrics.

\subsection{Comparing the uses of SALEDI and SAIDI}

SAIDI without major event days and SALEDI quantify different aspects of CMIp  and are complementary.
SAIDI without major event days focuses on the CMIp of common individual outages and quantifies reliability by excluding major event days, whereas SALEDI focuses on the CMIp of large events and quantifies resilience.

\looseness=-1
SAIDI with major event days measures the total impact of all sustained outages experienced by customers in a year.
Although SAIDI with major event days is some sort of combination of reliability and resilience, it varies erratically as it is mostly driven by the very large variability of the large events that can occur in each year. 
Moreover, it is desirable to separately monitor reliability and 
 resilience with different metrics. While improvements to reliability can often also improve resilience or  not affect resilience, they are distinct because the scale and urgency and risk of large events differ so much from routine outages. Therefore proposed upgrades should be evaluated separately for reliability and 
 resilience.

\subsection{CVAR interpretations}

We first review the definition of Conditional Value At Risk CVAR, writing $V$ for the value.
Suppose that $V$ has probability density function $f_V$.
We choose $V_{\rm large}$ at quantile $\beta$. That is, the probability that an event is large ($V\ge V_{\rm large}$) is $1-\beta$. $V_{\rm large}$ is the Value At Risk or VAR for quantile $\beta$. 
Also write $V_{\ge}$ for $V$ conditioned on $V\ge V_{\rm large}$. Then $f_{V_{\ge}}(v)=f_{V}(v)/(1-\beta)$ for $v\ge V_{\rm large}$.
Then CVAR is defined as\footnote{There can be an upper limit $V_{\rm max}$ in the distribution $f$  so that the integration can have upper limit $V_{\rm max}$.}
\begin{align}
{\rm CVAR}&=\frac{1}{1-\beta}\int_{V_{\rm large}}^\infty vf_V(v) dv=\int_{V_{\rm large}}^\infty vf_{V_{\ge}}(v) dv
\label{CVARF}
\end{align}
For an empirical pdf based on observations $V_1,V_2,...,V_{n_{\rm large}}$ that are $\ge V_{\rm large}$, each observation has probability $1/n_{\rm large}$ and we have 
\begin{align}
    f_{V_{\ge}}(v)&=
\sum_{i=1}^{n_{\rm large}} \frac{1}{n_{\rm large}}\delta(v- V_i)
\end{align}
where $\delta$ is a delta function, and substituting in \eqref{CVARF} gives
\begin{align}
{\rm CVAR}
&=
\frac{1}{n_{\rm large}}
\sum_{i=1}^{n_{\rm large}}  V_i
\label{CVARformula}
\end{align}

Now we can relate CVAR to other indices.
If the value is $ V=M/M_{\rm large}$, the normalized event CMIp,
(\ref{CVARformula}) becomes
\begin{align}
{\rm CVAR}_{\rm M}
&=
\frac{1}{n_{\rm large}}
\sum_{i=1}^{n_{\rm large}} M_i/M_{\rm large} ={\rm SPALED}
\label{CVARM}
\end{align}

If the value is $V=\ln (M/M_{\rm large})$, the logarithm of normalized event CMIp, (\ref{CVARformula}) becomes 
\begin{align}
{\rm CVAR}_{\rm lnM}&= \frac{1}{n_{\rm large}}
    \sum_{i=1}^{n_{\rm large}}\ln (M_i/M_{\rm large})= {\rm ALED}
    \label{CVARlnM}
\end{align}
Therefore ALED, which works well as an index, is a logarithmic form of CVAR\footnote{However, ${\rm CVAR}_{\rm lnM}$ differs from ALED in the choice of the threshold $M_{\rm large}$ for large events. 
For ${\rm CVAR}_{\rm lnM}$, $M_{\rm large}$ is VAR, the value at a given quantile of normalized event CMIp, whereas for ALED as presented here, $M_{\rm large}$ is the lowest point of the linear region of the exceedance function.}.
Moreover, (\ref{SALEDIfactor}) shows that 
 \begin{align}
{\rm SALEDI} = f_{\rm large} {\rm CVAR}_{\rm lnM}
\end{align}
so that one can regard SALEDI as an annual logarithmic form of CVAR that also accounts for the frequency of large events.

If customer cost $C$ of power outage for the average customer is assumed proportional to CMIp so that  $C=k M$ and the value 
 is $V=\log_{10} C$, then (\ref{CVARformula}) becomes 
 \begin{align}
{\rm CVAR}_{\rm logC}&= \frac{1}{n_{\rm large}}
    \sum_{i=1}^{n_{\rm large}}\log_{10} C_i={\rm ALEC},
     \label{CVARlnC}
\end{align}
the Average Large Event Cost risk metric of \cite{AhmadPESL25}.

${\rm CVAR}_{\rm M}$ in \eqref{CVARM} has all the high variability and statistical problems of SPALED discussed in section \ref{whylog}, making it impractical for tracking resilience with utility data.
Whereas ${\rm CVAR}_{\rm lnM}={\rm ALED}$ in \eqref{CVARlnM} gives a practical index. 
The logarithmic transformation also works for ${\rm CVAR}_{\rm logC}={\rm ALEC}$ in \eqref{CVARlnC}.
Thus CVAR works well as a resilience index only with a logarithmic value.

\subsection{Heavy tails and extrapolating the trend of extreme events}
\label{extrapolate}

\looseness=-1
The bounded Pareto distribution (\ref{Pb}) 
linearly extrapolates the slope of the exceedance curve tail beyond the maximum observed normalized CMIp $M_{\rm maxobs}/M_{\rm large}$ up to the maximum normalized CMIp
 $M_{\rm max}/M_{\rm large}$ for the largest possible blackout.
This allows the resilience to be evaluated while including the unobserved future blackouts of the highest customer impact. 
The linear extrapolation of the trend is somewhat uncertain, but the extrapolation nevertheless seems more reasonable than ignoring the even more extreme events indicated by the observed trend---it seems unlikely that the largest possible blackout has already been observed in the data. 
It is the case that sums of samples of heavy-tailed distributions are dominated by the maximum sample that occurs\footnote{This is called a ``catastrophe principle" and proved in \cite[chapter 3]{Nairbook22}.}. For example, both SAIDI with major event days and SPLEDI add up normalized CMIp, and suffer high variability due to this effect. 
High-impact blackouts with extremely high CMIp values are rare but inevitable due to heavy tails. 
This emphasizes the importance, despite its inherent uncertainties, of trying to estimate the extrapolated trend governing the most extreme events to encompass these extreme events in the resilience index. 
The need to obtain this linear extrapolation is one reason that we confine the large events to the linear region of the tail. We assume the existence of a suitable linear region across a sufficient range of the largest observed blackouts because it is present in the data we analyze here.

\section{Practical details of SALEDI}\label{sec:PracticalDetails}

\looseness=-1
This section gives guidelines for selecting the parameters needed to compute SALEDI.

\subsection{Choosing the large event threshold $M_{\rm large}$}\label{sec:ChoosingMlarge}
\label{threshold}
\begin{figure}[ht]
    \centering
    \includegraphics[width=1.0\linewidth]{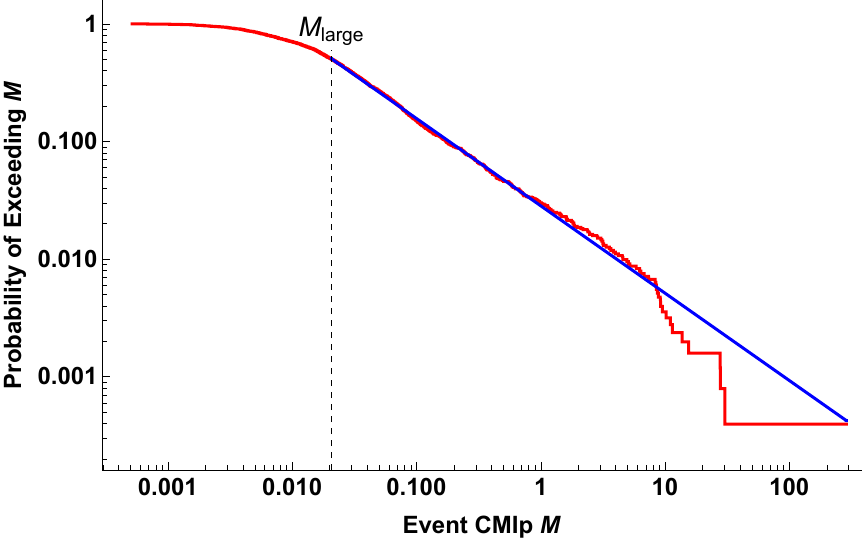}
    \caption{Probability exceedance function of event CMIp of Utility 2, overlaid with  $M_{\rm large}$ and the corresponding estimated straight line Pareto distribution.}
    \label{fig:MlargeExample}
\end{figure}

SALEDI is defined over events whose CMIp $M_i$ exceeds the $M_{\rm large}$ threshold. It is intended that $M_{\rm large}$ is calculated once and then stays fixed thereafter. 
We want to select an $M_{\rm large}$ that satisfies the following criteria:
\begin{itemize}
    \item $M_{\rm large}$ should be small enough to give sufficient large events to keep the variability ${\rm RSE}_{\rm SAL}$ sufficiently small.
    \item $M_{\rm large}$ should be within the approximately linear region (on a log-log plot) of the tail of the distribution of $M_i$. 
\end{itemize}
To satisfy the above criterion, we select $M_{\rm large}$ using a statistically principled procedure proposed by Clauset et al. \cite[section 3.3]{ClausetSIAM09} and summarized as follows:

Let $\{M_i\}$ denote all values of event CMIp across all observed years. Candidate thresholds are formed from upper-tail values of $\{M_i\}$.
For each candidate threshold $M_0$, we repeat the following steps:
\begin{enumerate}
    \item Select all values above $M_0$, i.e., $S_0=\{M_i:M_i \geq M_0\}$
    \item Estimate $\alpha_0$ using the selected data:\\
    $\alpha_0=(1/|S_0|)\sum_{i=1}^{|S_0|}{\rm ln} \{M_i/M_0:M_i \geq M_0\}$\\
    where $|S_0|$ is the number of values in the selected data.
    \item Calculate Kolmogorov-Smirnov distance $D$ between the empirical CDF of the selected data $F_{S_0}(M)$ and the CDF $F_{P_0}(M)$ of a Pareto distribution with parameter $\alpha_0$:
    $D=\underset{M_i \geq M_0}{\rm max} |F_{S_0}(M_i)-F_{P_0}(M_i)|$
\end{enumerate}
The $M_0$ value that yields the minimum Kolmogorov-Smirnov distance $D$ among all candidate thresholds is selected as $M_{\rm large}$. This procedure identifies the smallest threshold for which the tail behavior is well approximated by a power law, thereby avoiding both contamination from non-tail events (if $M_{\rm large}$ is too small) and unnecessary loss of accuracy (if $M_{\rm large}$ is too large). 
The $M_{\rm large}$ value calculated using this method is shown on the probability exceedance function of event CMIp in Fig.~\ref{fig:MlargeExample}. The $M_{\rm large}$ values for the five utilities and their corresponding quantiles $q$ are given in Table~\ref{utilitydata}.

\subsection{Number of large events  $n_{\rm large}^{\rm min}$ required for accuracy}
\label{accuracy}
The statistical variability of SALEDI is governed by the number of large events $ n_{\rm large}$ used, as can be seen from \eqref{RSESALEDI}. 
Hence, rearranging \eqref{RSESALEDI},
to ensure for SALEDI a statistical accuracy ${\rm RSE}_{\rm SAL}^{\rm max}$ so that ${\rm RSE}_{\rm SAL}\le{\rm RSE}_{\rm SAL}^{\rm max}$, the minimum number of large events required is
\begin{align}
 n_{\rm large}^{\rm min}&= 
2/({\rm RSE}_{\rm SAL}^{\rm max})^2
    \label{nlarge}
\end{align}

Suppose that $n_{\rm large}^{\rm all}$ large events are observed in all the outage data over $n_{\rm year}^{\rm all}$ years.
Then the average annual frequency of large events using all the observed years of data is
\begin{align}
    f_{\rm large}^{\rm all}=n_{\rm large}^{\rm all}/n_{\rm year}^{\rm all}
\end{align}
and the minimum number of years of data  required to ensure for SALEDI a statistical accuracy ${\rm RSE}_{\rm SAL}^{\rm max}$ is
\begin{align}
 n_{\rm year}^{\rm min}=  
 n_{\rm large}^{\rm min}/f_{\rm large}^{\rm all}
    \label{RSEcalc2}
\end{align}

In this paper, we adopt ${\rm RSE}_{\rm SAL}^{\rm max}= 0.1$, implying that the standard deviation of an estimate of SALEDI is no more than 10\% of its mean value.
Then \eqref{nlarge} gives   $n_{\rm large}^{\rm min}= 200$.

Under reasonable normality assumptions, ${\rm RSE}_{\rm SAL}^{\rm max}= 0.1$ further indicates that $\sim$68\% of the estimates of SALEDI lie within 10\% of the mean, and that $\sim$90\% of the estimates of SALEDI lie within 16\% of the mean.

The $n_{\rm year}$  values for the five utilities based on ${\rm RSE}_{\rm SAL}^{\rm max}= 0.1$ are given in Table~\ref{utilitydata}. 
The value of $n_{\rm year}^{\rm min}$ implies tracking SALEDI with an $n_{\rm year}^{\rm min}$ year sliding window. This is practical for the five  utilities in Table~\ref{utilitydata} that have $n_{\rm year}^{\rm min}\approx$ 2, 3, or 5.

\section{Conclusions}

Quantifying distribution system resilience in terms of the observed customer minutes interrupted per customer served (CMIp) in large events can be considered one of the basic challenges for resilience metrics.
However, detailed outage data from 5 utilities in the USA shows heavy-tailed distributions and high variability in 
event CMIp. Indeed these observed data show evidence of very heavy tails (slope magnitude $\alpha<1.5$ of an exceedance function on a log-log plot).
A consequence of the heavy tails is that simply adding 
up or averaging the CMIp for large events does not give practical indices. These impractical indices (such as SPALEDI, SPALED, and conventional CVAR) vary erratically and their
relative standard errors are usually so large that too many (one or two orders of magnitude more) observed large events would be needed to enable usable statistical accuracy. 

This paper solves this problem of heavy tails with a logarithmic transformation of the data to obtain the new resilience metric SALEDI describing the annual customer impact of large events. To calculate SALEDI, one adds the logarithms of the normalized large-event CMIp and then divides by the number of years observed. The number of observed years required for reasonable statistical accuracy is 2 to 5 years across the five utilities, making it practical to monitor and track resilience using standard utility data.
SALEDI could also be used as a metric for outage data produced by simulations, where it would require running substantially fewer cases.

\looseness=-1
SALEDI is the product of the annual frequency of large events $f_{\rm large}$ and the ALED metric that is an average magnitude of large events. 
More precisely, ALED is the mean of the logarithms of normalized large event CMIp. 
Risk is often described by exceedance functions, and resilience risk describes the tails of these functions.
ALED$^{-1}$ estimates the slope magnitude $\alpha$ of the tail of the probability exceedance function of CMIp on a log-log plot. 
Thus, ALED quantifies the trend of the observed large blackouts, which is significant because this trend may well govern future blackouts of high risk that are larger than those already observed. Our definition of large events is chosen so that the large events lie in the approximately linear tail region of the log-log plot to ensure that ALED, and in turn SALEDI, describes this trend. 
This approach requires a suitable approximately linear tail region for the largest blackouts, as observed in the data for the 5 utilities.
SALEDI is normalized by the number of customers served, so it is comparable across utilities of different sizes.

The paper also systematically interprets and compares the metrics SALEDI and ALED, which can be considered to be areas under the tails of  exceedance functions. ALED is a version of CVAR in which the value is the logarithm of normalized event CMIp. SAIDI is the area under a frequency exceedance function for outage CMIp. 

Resilience mitigation should decrease the frequency and severity of customer impacts, and in this regard,
$f_{\rm large}$, ALED,  and SALEDI  usefully measure different aspects.
$f_{\rm large}$ is the frequency of large events, which is driven by the extremes of weather exceeding a given customer impact.
ALED measures an average (on a log scale) large event impact for the customers, which results from a combination of weather severity,  infrastructure hardness, and restoration speed. 
A utility with high $f_{\rm large}$ but low ALED is being hit frequently but withstands and recovers well relative to the weather severities. 
SALEDI combines ALED and $f_{\rm large}$ to measure the total annual large event customer impact.
SALEDI gives a single value for optimization and tracking resilience over the years, that can play for resilience for customers the same role as SAIDI does for reliability.
A utility would first compute the large event threshold $M_{\rm large}$ and the required $n_{\rm year}^{\rm min}$ once from its full historical dataset. 
It would then maintain a rolling $n_{\rm year}^{\rm min}$ window and update SALEDI annually to track resilience.

\looseness=-1
In summary, since the CMIps in large events have high variability and heavy-tailed distributions, simply adding up or taking the mean of CMIp in large events does not produce practical indices. A logarithmic transformation of CMIp converts heavy tails to light tails for which the usual statistics and indices work. Thus an overall conclusion is simple: Based on five cases of distribution utility data in the USA, CMIp in large events should be measured on a logarithmic scale, so that practical resilience indices tracking customer minutes interrupted in large events should be calculated using the logarithm of the CMIp. 

\printbibliography

\appendices

\section{}\label{AppendixA}
The plausibility of a Pareto tail (\ref{powerlaw}) is not rejected for utilities 1,2,5 at a p-value of 0.1, using Clauset's goodness-of-fit test \cite[Section 4]{ClausetSIAM09}.
The plausibility of a lower-bounded truncated lognormal fit\footnote{
The truncated distribution is a lognormal distribution conditioned on $\left[a,\infty\right)$ with $a>0$ with exceedance function  
$\overline{F}_{Y}(p)=
\overline{F}_N(\ln p)/
\overline{F}_N(\ln a)$
where $\overline{F}_N$ is the probability exceedance function of a normal distribution. $a$ is chosen using Clauset's method applied to a truncated lognormal.
} to the tails is not rejected at a p-value of 0.1 for utilities 1,2,3,4. 
A likelihood ratio test comparing Pareto and lognormal distributions shows insufficient evidence to prefer one over the other at a p-value of 0.1.

The truncated lognormal fit is quadratic on a log-log plot, but with a large enough $\sigma$ so that the quadratic coefficient $-1/(2\sigma^2)$ is small \cite[section 1.2.3]{Nairbook22} and the linear approximation is acceptable.
The standard deviation parameter $\sigma$ of the best-fit truncated lognormal distribution for the utilities 1,2,3,4,5 are 4.05, 4.53, 3.80, 4.16, 1.39, respectively.

Formal testing of the tails to determine whether the data is likely to have been drawn from particular statistical distributions is relevant. 
However, it is not the only consideration, and formal testing does not directly answer whether the linear approximation of the tails is a close enough approximation for constructing an index.
While both linear and quadratic approximations are reasonable approximations, and indeed statistically indistinguishable, we choose the linear approximation to construct SALEDI for several other reasons.
The linear approximation is simpler and is conservative compared to the mildly concave quadratic approximation obtained by fitting the log-normal distribution. 
Moreover, the linear approximation uses the single parameter $\alpha$ to describe the tail, producing a single index. 
The quadratic approximation describes the tail with two parameters and would have produced two indices.

\section{}\label{AppendixB}
Let $N_{\rm large}$ be  Poisson[$ n_{\rm large}$], assumed independent of i.i.d. samples $X_1$, $X_2$, ..., $X_{N_{\rm large}}$ of random variable $X$. Let
\begin{align}
    S=\frac{1}{n_{\rm year}}
    \sum_{i=1}^{N_{\rm large}}X_i
   \quad\text{ and }\quad
     S_{\rm mean}=\frac{1}{N_{\rm large}}
    \sum_{i=1}^{N_{\rm large}}X_i 
    \notag
\end{align}
${\rm E} N_{\rm large}= {\rm Var}N_{\rm large}= n_{\rm large}$. A Wald equation gives ${\rm E}S=
 n_{\rm large}{\rm E}X/n_{\rm year}$ 
and a Blackwell-Girshick equation gives 
\begin{align}
    ~\hspace{-5mm}{\rm \hspace{-5mm}Var}S
    = \,&({\rm E}[N_{\rm large}]{\rm Var}X+{\rm Var}[N_{\rm large}]({\rm E}X)^2)/n_{\rm year}^2
    \notag\\
    = \,& n_{\rm large}({\rm Var}X+({\rm E}X)^2)/n_{\rm year}^2
    \text{ and}
    \notag\\
    {\rm RSE}_S
    &=\frac{\sigma[S]}{{\rm E}S}
    =\frac{\sqrt{1+{\rm Var}X/({\rm E}X)^2}}{\sqrt{ n_{\rm large}}}
    =\frac{\sqrt{1+({\rm RSE}_X)^2}}{\sqrt{ n_{\rm large}}}\notag
    \end{align}
Also
${\rm E}S_{\rm mean}={\rm E}X$ and
${\rm Var}S_{\rm mean}
    =  ({\rm Var}X)/ n_{\rm large}$ and 
\begin{align}
  {\rm RSE}_{S_{\rm mean}}&=  \frac{\sigma[S_{\rm mean}]}{{\rm E}S_{\rm mean}}
  = \frac{\sigma[X]}{{\rm E}X\sqrt{ n_{\rm large}}}
  =\frac{{\rm RSE}_X}{\sqrt{ n_{\rm large}}}.\notag
\end{align}
To calculate ${\rm RSE}_{Pb}$ and ${\rm RSE}_{LNb}$  we use 
${\rm RSE}_{Y} = \sqrt{{\rm E}[Y^2]/{\rm E}[Y]^2-1}$ and, for $\alpha \neq 1,2$, and $k=1$ or $k=2$,
\begin{align}
    {\rm E}[(Pb)^k] &= \frac{\alpha}{\alpha - k}\frac{1-(p_{\rm max})^{k-\alpha}}{1-(p_{\rm max})^{-\alpha}}  \notag\\
    {\rm E}[(LNb)^k] &= e^{k\mu + \frac{k^2\sigma^2}{2}}\  \frac{\Phi(\frac{{\rm ln}p_{\rm max}-\mu -k\sigma^2}{\sigma}) - \Phi(\frac{-\mu -k\sigma^2}{\sigma})}{\Phi(\frac{{\rm ln}p_{\rm max}-\mu}{\sigma})-\Phi(\frac{-\mu}{\sigma})}\notag 
\end{align}

\end{document}